\documentclass[onecolumn]{aa}
\usepackage{graphicx}
\usepackage{txfonts}
\def\AA {A\&A{ }}
\def\ApJ {ApJ{ }}
\begin{document}
\title{The Coma Cluster hard X-ray spectrum revisited: still no evidence for a hard tail.}
\titlerunning{The Coma Cluster hard X-ray spectrum revisited}
\author{Mariachiara Rossetti 
\inst{1}
\and 
 Silvano Molendi
\inst{1}}
\institute{
          IASF-Milano, INAF, via Bassini 15
         I-20133 Milano, Italy
	 \email{rossetti@iasf-milano.inaf.it}}
\authorrunning{Rossetti \& Molendi}

%\title{The Coma Cluster hard X-ray spectrum revisited: delusions die hard.}
%\maketitle
\date{\today}
\abstract{
In this note, we reply to Fusco-Femiano et al. 2004 (FF04) and  Fusco-Femiano et al. 2006 (FF06), who cast doubts on our analysis of the PDS observations of the Coma Cluster which we describe in Rossetti \& Molendi 2004 (RM04). We discuss the main issues in FF06 and we confirm that the available data do not allow to firmly establish the presence of a non-thermal component in the hard X-ray spectrum of the Coma cluster.
}
\maketitle
\section{Introduction}   
Non thermal emission in the X-ray band is predicted in cluster of galaxies showing extended radio emission, since the relativistic electrons can interact with the CMB photons and produce Inverse Compton X-ray radiation. The first clear detection of this component has been found in the spectrum of the Coma Cluster observed with the PDS instrument on board {\it BeppoSAX} by Fusco-Femiano et al. 1999 (F99), who found an excess above the thermal emission significant at $4.5\sigma$. However, new analysis of this observation by RM04 and FF04 have identified a trivial error in the spectra extraction lowering the statistical significance of the excess to $2.5\sigma$ and $3.4\sigma$ respectively. \\
FF06 discuss the origin of the discrepancy in the results, as well as the differences  in the results of the analysis of a second PDS observation (RM04 report an excess which is significant at less than one $\sigma$ while FF04 find an excess at $3.4 \sigma$). FF06 suggest that these differences could be due to the different software for data reduction and to a different treatment of the background.\\
In this note, we discuss the issue of the background subtraction, we underline some important assumptions on which the results are based and we confirm our previous results.
 
\section{Background subtraction}
The PDS instrument uses the rocking collimator technique to subtract the background. The standard observation strategy is to observe the target with one collimator while the other monitors the background and to periodically swap them. Two offset positions are available and they are both used in the standard background subtraction. A careful check for contaminating sources in the OFF fields should be performed.\\
FF04 and RM04 both report an excess of count rates in the $OFF^{+}$ position, with respect to the $OFF^{-}$, in the second observation of the Coma cluster.
FF04 suggest that this difference is due to a variable contaminating source in the $OFF^{+}$ field and therefore use only the $OFF^{-}$ field as background. In our previous paper (RM04), we argued that this difference could be only a statistical fluctuations, by performing a detailed analysis of the differences between the OFF fields in a sample of 69 long PDS observations at high galactic latitudes (Section $2.1$ of RM04). We found that there is a systematic difference, probably of instrumental origin, between the OFF fields: $<OFF^{-}-OFF^{+}>=(2.44\pm0.33)\times10^{-2}$ cts/s in the 25-80 keV energy range, while the dispersion around the mean is large $\sigma=4.33 \times10^{-2}$ cts/s. This dispersion is not due to the ``small'' number of observations considered, the width is intrinsic to the distribution (as confirmed by a maximum likelihood test) and a significant contribution is likely given by the fluctuations of the Cosmic X-ray background. The difference between the $OFF$ fields observed in the second Coma observation is consistent with being a statistical fluctuation at $1.1\sigma$. A systematic difference between the OFF fields is also reported  by Nevalainen et al. (2004), who suggest that a correction factor should be added (with the proper sign) when only one of the background fields is considered. FF06 find that there is no systematic difference between the OFF fields but, since they do not provide details of this analysis, we could not compare our results.\\
In RM04 we presented in detail results obtained using both background fields, but we also showed that using only the $OFF^{-}$ fields spectra are consistent and we could find an excess over the thermal emission significant only at $1.7 \sigma$ (Section 3.2 in RM04).\\
%Therefore the sentence in Sec. 3 of FF06 (``the check regarding the presence of contaminating sources in the +OFF field was not performed by RM04'') is not true. \\
Another background component that could partly affect the results is due to the charged particles that interact with the detectors, which produce spurious spikes in the light curve of the observation. In our analysis, we applied a $\sigma-$clipping technique to reject spurious spikes  and therefore we exclude that this kind of contaminating events could have biased our spectra. Unfortunately, we forgot to explicitly state this in RM04. \\
%Unfortunately, we did not write in RM04 that spike filtering was applied to our data with a $\sigma-$clipping technique and therefore we exclude that this kind of contaminating events could have biased our spectra.    

\section{Temperature value}
The excess above the thermal emission is calculated by fitting the spectrum with a bremsstrahlung model with the temperature fixed at a value found in the literature and by comparing the observed count rate and its error with the count rate predicted by the model in a given energy band:
\begin{equation}
\rm{excess}=\frac{observed-predicted}{error}.
\label{eq_eccesso}
\end{equation}   
In RM04, we reported the discrepancy in the temperature values measured by various instruments and we argued that the temperature should be allowed to vary between 8 and 9 keV. While comparing our results with those in FF99 we used the \emph{Ginga} value $8.21 \pm 0.16$ keV (Hughes et al. 1993). In FF04 and FF06 the temperature is fixed to a lower, although consistent,  \emph{Ginga} value: $8.1$ keV (David et al. 1993). The main problem of the analysis presented in FF99, FF04 and FF06 is that the authors fix the temperature at a best fit value, without even allowing this parameter to vary between its $1\sigma$ errors. 
Indeed, the temperature value is a crucial parameter in determining the excess and the dependence of the results on the temperature assumption should be studied in detail.\\
\begin{figure*}[!tp]
 \centering
\vbox{
   \includegraphics[width=0.7\textwidth, angle=0]{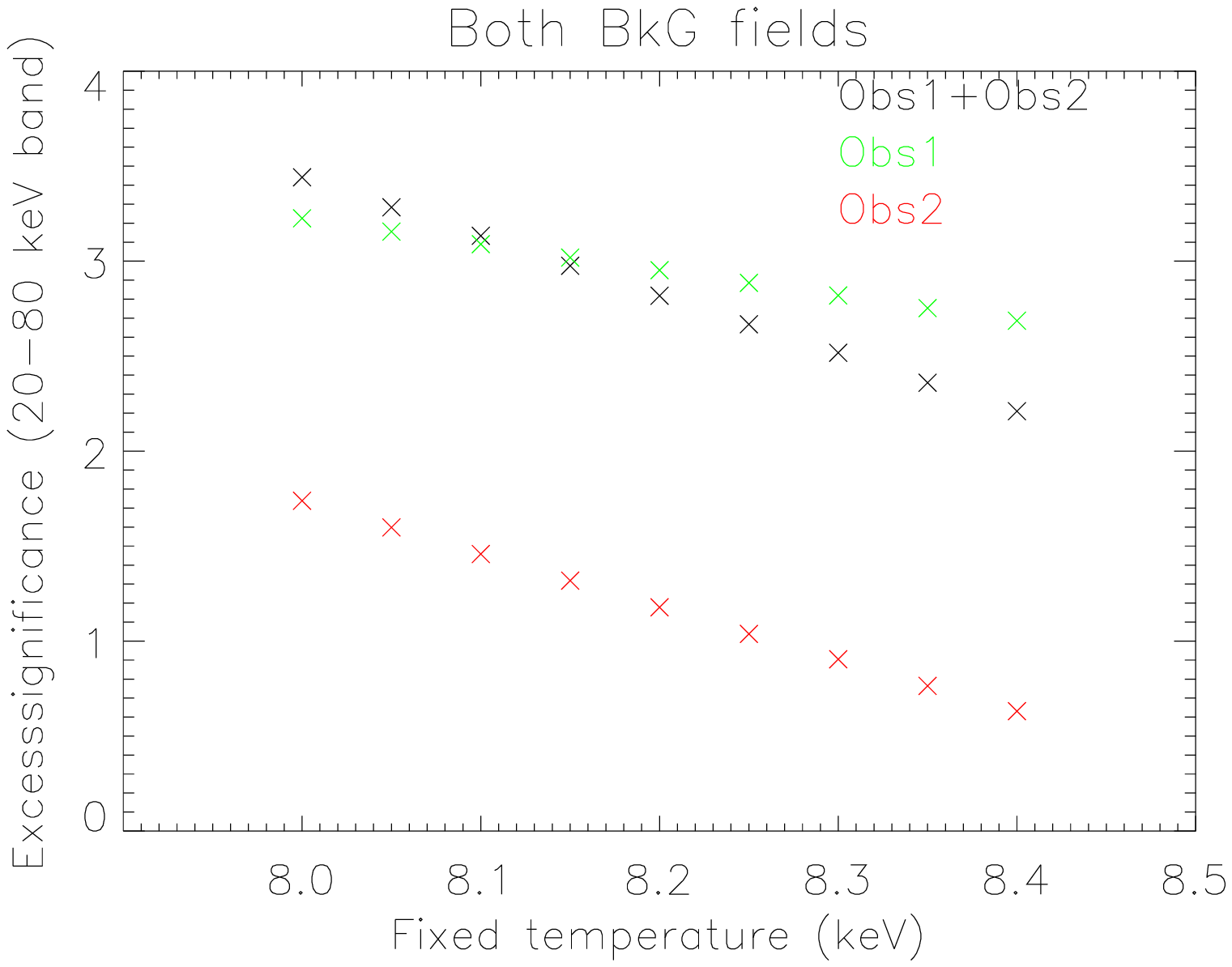}
    \includegraphics[width=0.7\textwidth, angle=0]{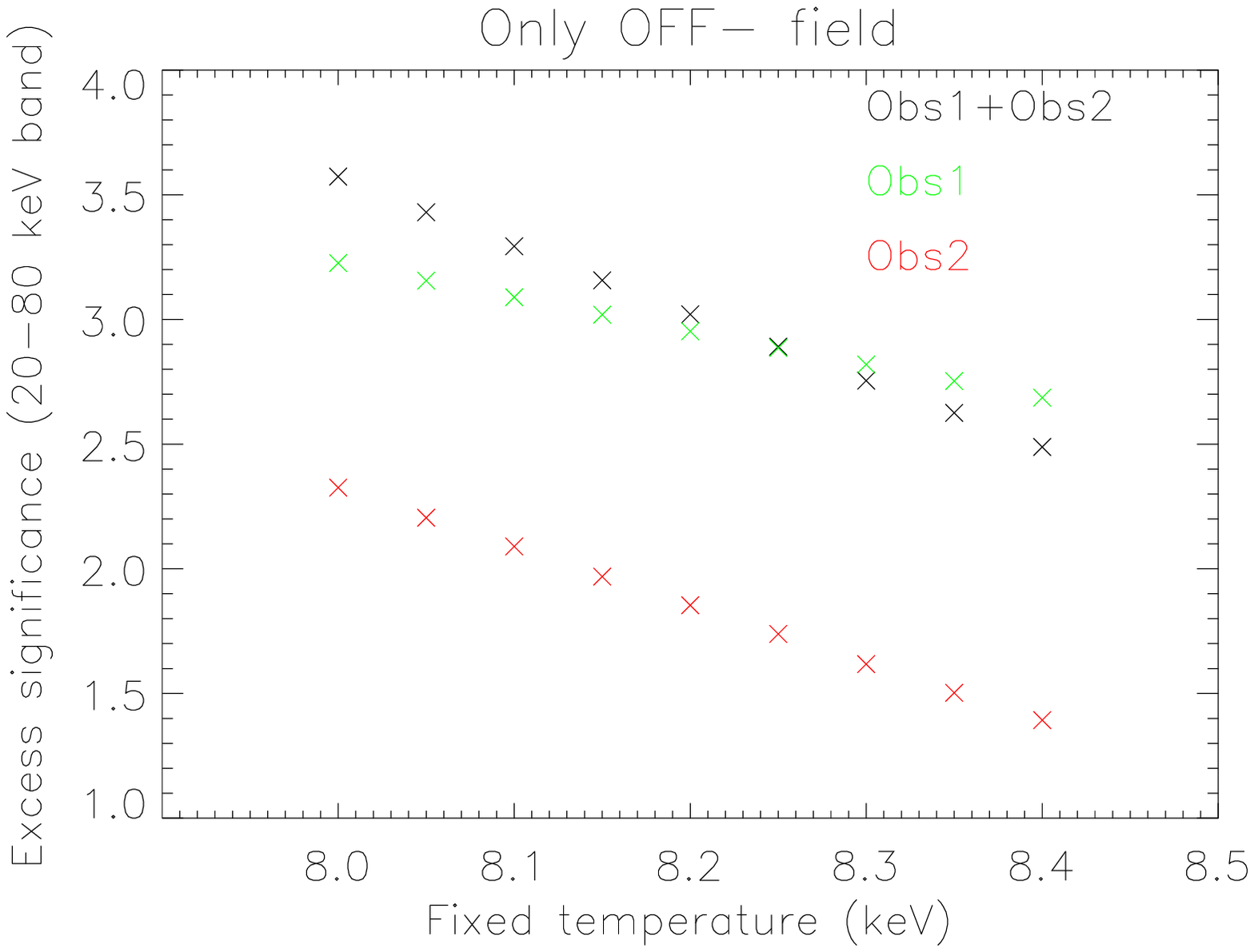}
}

%\label{fig_eccessi}
\caption{\label{fig_eccessi}\emph{Upper panel}: Excess significance (calculated as in Eq. \ref{eq_eccesso} in terms of $\sigma$) in the energy band 20-80 keV using both background fields. The black points are the results for the sum of the observations, green for the first observation and red for the second ; \emph{Lower panel}: the same as before but using only the $OFF^-$ field as background. The black points can be compared directly with the results in FF06}
\end{figure*}
To study this dependence, we have performed an exercise varying the  temperature in the range $8-8.4$ keV (i.e. within the $90\%$ c.l. error bars of the \emph{Ginga} measurement $8.21 \pm 0.16$ keV) and calculating the corresponding excess. In Figure \ref{fig_eccessi} we show the significance of the excess for the Coma observations and for the sum of their spectra as a function of the temperature, using both OFF fields as background in the upper panel and only the $OFF^-$ field in the lower panel.  These figures show that the significance of the excess decreases rapidly and falls below $3\sigma$ as the temperature increases and that there are some differences between the observations (we will discuss this issue in the next section).\\
This exercise shows that the temperature is a crucial parameter in determining the statistical significance of a non thermal excess. Given the large discrepancies in the temperature estimate of the Coma Cluster (see RM04 and references therein), a large temperature interval including the values found by other instrument than \emph{Ginga} should be considered, as in RM04.\\ 
%Moreover, given the large discrepancies in the temperature estimate of the Coma Cluster (see RM04 and references therein), a larger temperature interval including the values found by other instrument than Ginga should be considered.\\ 
It is interesting to note that the results reported by FF06 is indeed one of the ``best'' results available under reasonable assumptions: the choice of using the sum of the observations, the $OFF^-$ field as the background and a fixed temperature of $8.1$ keV allows to obtain one of the ``largest'' excess over the thermal emission, but also the results obtained under different assumptions should be considered. 

\section{Differences between the observations} 
\begin{figure*}[!tbp]
\centering
\includegraphics[width=0.7\textwidth, angle=0]{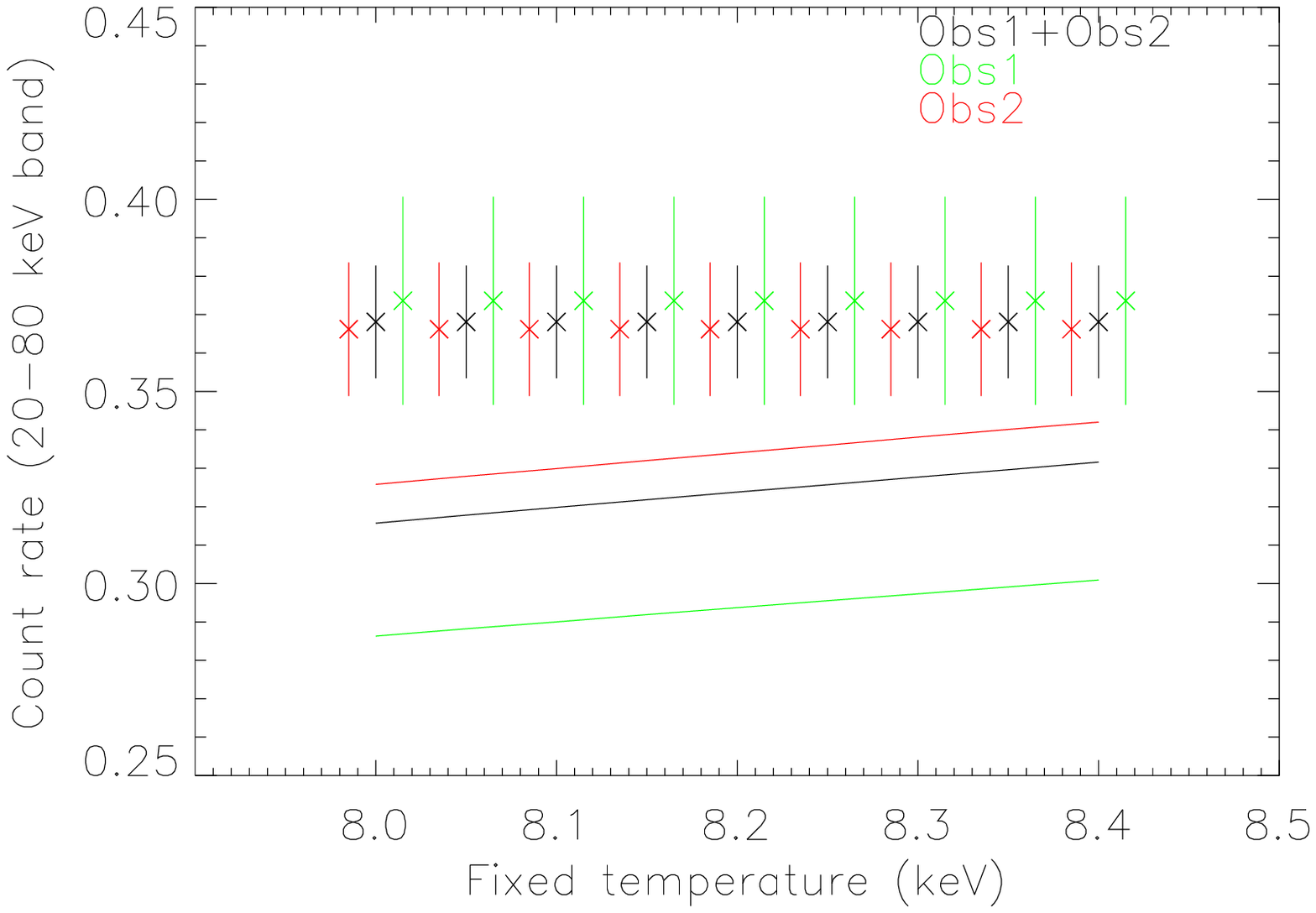}
%\psfig{figure=fig2.ps,angle=0, width=\textwidth}%
%\label{fig_crate}
\caption{\label{fig_crate}Observed (crosses) and predicted (lines) count rates in the 20-80 keV energy band obtained using the -OFF field as background. Black symbols are for the sum of the observations, green for the first and red for the second.}
\end{figure*}
It is apparent in Figure \ref{fig_eccessi}, that different results on the excess estimate can be obtained from the available observations: the excess measured with the first observation is much larger than that obtained with the second.  In Figure \ref{fig_crate} we compare the observed count rates, and their errors, with the predicted values from the fitting model at a given temperature. It is interesting to note that the measured count rates are consistent within the errors and that the error bar reduces only slightly in the sum of the observations with respect to the second alone (this is expected since the exposure time of the second observation is much larger than the first one and since the background level of the second is lower than that of the first, due to the lowering of the orbit, as discussed in RM04 and FF06). Therefore the difference between the excess  measured in the second observation and in the sum (Figure \ref{fig_eccessi}) cannot be due to the size of the errors. Indeed Figure \ref{fig_crate} shows clearly that the differences are in the predicted values and therefore they are due to the fitting models, in particular to the only free parameter of the fit: the normalization. For instance, fixing the temperature to 8.1 keV, the normalization values for observation 1 and 2 are inconsistent at $4.9\sigma$, leading to predicted count rates 0.29 for the first observation and 0.33 for the second, which are significantly different.\\
\begin{figure*}[!tb]
\centering
\includegraphics[width=0.5\textwidth, angle=270]{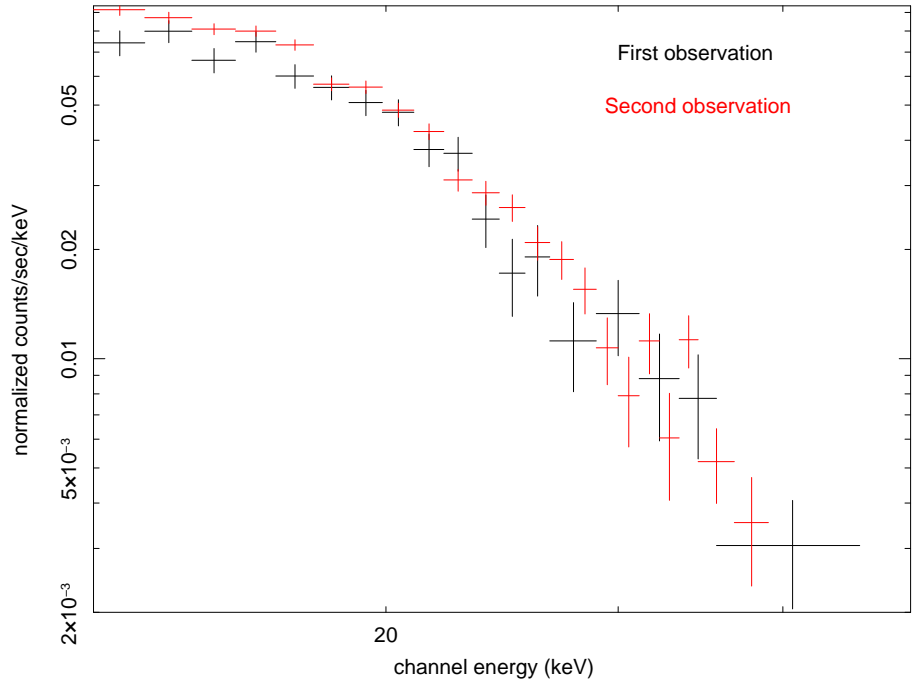}
%\psfig{figure=fig3.ps,angle=270, width=\textwidth}%
%\label{fig_confro}
\caption{\label{fig_confro} Spectra of the first (black) and second (red) Coma observation, using both background fields. They are significantly different below 20 keV.}
\end{figure*}
The two observations are not significantly different in the 20-100 keV energy range, but the differences are significant below 20 keV (Figure \ref{fig_confro}). This could explain the differences between the predicted count rates, which indeed reduces by eliminating channels below 20 keV. The origin of this effect is not clear: it is not affected by the choice of the background file and therefore it is related to something in the ON field.  However, eliminating the low energy channels (up to 20 keV) during the fitting procedure the excess significance in both observations is even lower than with the fit in the whole energy range.\\
The choice of using the summed spectra of the observations could be justified in order to improve the statistics, but since the second observation is three times longer than the first and the intensity of the background is lower \footnote{As already outlined in RM04 and FF06, the intensity of the background during the second observation was $20\, \%$ lower than during the first, due the {\it BeppoSAX } orbital decay}, the statistical error with the second observation alone is only slightly larger than with the sum (Figure  \ref{fig_crate}). Moreover, this choice introduces a systematic uncertainty, not easy to estimate, due to the differences between the observations. Therefore a small improvement in the statistics is obtained at the price of introducing an unknown systematic error.   This is why it is better to consider the observations separately and not use the summed spectrum.\\ 
%However Figure  \ref{fig_crate} shows that the statistical error with the second observation alone is only slightly larger than with the sum. Although slightly reducing the statistical error, this choice introduces a systematic uncertainty, not easy to estimate, due to the differences between the observations. This is why it is better to consider the observations separately and not use the summed spectrum.\\
% Moreover the second observation is ``better'' than the first because, as already outlined in RM04 and FF06, the intensity of the background during the second observation was $20\, \%$ lower than during the first, due the {\it BeppoSAX } orbital decay. 

\section{Conclusions}
In this note, we have shown that the measurement of the excess above the thermal emission in the PDS observation of galaxy clusters is based on several assumptions, namely:
\begin{itemize}
\item {the choice of the background, since two OFF fields are available;}\\
\item {the choice of the temperature value for the thermal model;}\\
\item {the choice of the observation, the two spectra separately or their sum.}
\end{itemize}  
Indeed, FF04 and FF06 have shown that under a given set of assumptions (a source contaminating the $OFF^+$ field, a global temperature of $8.1$ keV) a non thermal excess can be found in the PDS observations of the Coma cluster. However, their assumptions are invariably those leading to a larger excess.   We have shown that for other choices of the critical parameters, some of which more appropriate, the significance of the excess is lower. \\
%always correspond to the choices that allow to measure the  largest excess.   We have shown that other reasonable choices of the critical parameters, the significance of the excess can be lowered. \\
The heart of the matter is that all "reasonable" choices should 
be taken  into acount when establishing the statistical significance of 
the excess.
%This is indeed one of the most elementary rules in ...
Failure to do so, as in the case at hand, will almost invariably result in 
a spurious detection. 
We conclude that  available data do not allow to firmly establish the presence of a non-thermal component in the hard X-rays spectrum of the Coma cluster.
% A fundamental results as the detection of hard X-ray emission from clusters of galaxies should not be based on one single combination of these parameters, but all the different cases should be discussed. From this analysis we conclude that  available data do not allow to firmly establish the presence of a non-thermal component in the hard X-rays spectrum of the Coma cluster. We have to wait until \emph{SUZAKU} or future instruments have the necessary sensitivity to detect this interesting component in the spectra of galaxy clusters.

\end{document}